\documentclass[11pt,twoside]{article}

\usepackage{asp2014}

\aspSuppressVolSlug
\resetcounters

\begin{document}
\title{Using Felis to Represent the Semantics and Metadata of Astronomical Data Catalogs}

\author{Jeremy~McCormick,$^1$ Gregory~P.~Dubois-Felsmann,$^2$ Andrei~Salnikov,$^1$ Brian~Van~Klaveren,$^1$ and Tim~Jenness$^3$}
\affil{$^1$SLAC National Accelerator Laboratory,  2575 Sand Hill Rd., Menlo Park, CA 94025, USA}
\affil{$^2$Caltech/IPAC, California Institute of Technology, MS 100-22, Pasadena, CA 91125-2200, USA}
\affil{$^3$Vera C.\ Rubin Observatory Project Office, 950 N.\ Cherry Ave., Tucson, AZ  85719, USA}
\paperauthor{Jeremy~McCormick}{}{}{SLAC National Accelerator Laboratory}{}{Menlo Park}{CA}{94025}{USA}
\paperauthor{Gregory~P.~Dubois-Felsmann}{}{0000-0003-1598-6979}{Caltech/IPAC}{}{Pasadena}{CA}{91125-2200}{USA}
\paperauthor{Andrei~Salnikov}{}{0000-0002-3623-0161}{SLAC National Accelerator Laboratory}{}{Menlo Park}{CA}{94025}{USA}
\paperauthor{Brian~Van~Klaveren}{}{}{SLAC National Accelerator Laboratory}{}{Menlo Park}{CA}{94025}{USA}
\paperauthor{Tim~Jenness}{}{0000-0001-5982-167X}{Vera C.\ Rubin Observatory Project Office}{}{Tucson}{AZ}{85719}{USA}

\hypersetup{
    pdftitle={Using Felis to Represent the Semantics and Metadata of Astronomical Data Catalogs},
    pdfauthor={mccormickj},
    pdfkeywords={}
}


\begin{abstract}
    The Data Management team of the Vera C. Rubin Observatory has developed a data description language and toolset, Felis, for defining the semantics and metadata of its public-facing data catalogs. Felis uses a rich Pydantic data model for describing and validating catalog metadata, expressed as a human-readable and editable YAML format. Felis also provides a Python library and command line interface for working with these data models. The metadata is used to populate the TAP\_SCHEMA tables for the IVOA TAP services utilized by the Rubin Science Platform (RSP). Felis's current capabilities will be discussed along with some future plans.
\end{abstract}



\section{Introduction}

Tabular data catalogs are a fundamental part of modern astronomical research, and relational databases are a commonly used technology for providing access to them, typically via SQL queries.
Such systems rely on Data Definition Language (DDL) for defining the structure, or schema of the catalog, but DDL lacks the ability to define catalog metadata such as units of measurement or relationships between columns.
This metadata is critical for understanding and processing the data, as well as presenting it to users in a meaningful way.

The Data Management team at the Vera C.\ Rubin Observatory \citep{2019ApJ...873..111I} has developed a data format for defining the metadata for a particular data catalog or schema.
Felis\footnote{\url{https://github.com/lsst/felis}} reads and validates this format using the Pydantic Python library\footnote{\url{https://pydantic.dev}}.
The validated model is used to generate DDL statements or populate a TAP\_SCHEMA database, which can then be accessed through a Table Access Protocol (TAP) service endpoint \citep{2019ivoa.spec.0927D}.
Felis may also be used for a variety of additional tasks, such as generating documentation or validating schema updates within continuous integration (CI) processes.

\section{Schema Data Model}

\articlefigure{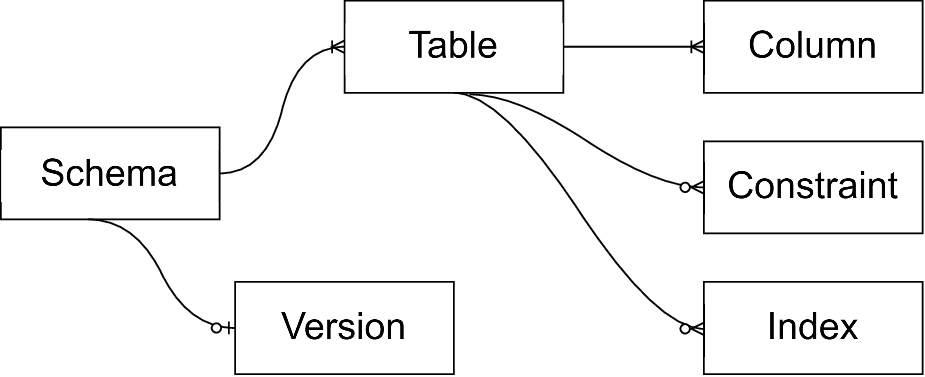}{fig:schema}{ERD diagram of the schema data model.}

All objects in the schema have a set of common attributes, including a name, identifier, and description.
The name attribute is required and corresponds to the name of the object in the target database.
The description is an optional, human-readable description of the object.
The identifier provides a way to uniquely identify the object within the schema and is primarily used for object referencing.
Identifiers may be generated automatically or provided explicitly by the user.

A version attribute indicates the version of the schema and can be used to track changes over time.
In addition to the version string, lists of compatible and read-compatible versions can also be provided.
Felis does not enforce any particular versioning scheme, but semantic versioning is recommended.

A schema must define at least one table, and each table must include a list of one or more columns.
Columns must have a \texttt{datatype} field and may include a number of optional properties which are used for a variety of purposes, including specifying semantics of the column, such as units or an IVOA Unified Content Descriptors (UCD) value \citep{2023ivoa.spec.0125C}.
These properties can also specify the column's behavior in a database, such as whether it is nullable.
Both tables and columns may define a TAP index, typically used to recommend an ordering for clients.

Constraints define rules that restrict column values in a table.
Felis supports primary key, foreign key, and unique constraints.
Each constraint defines the columns to which it applies, as well as the target primary key for a foreign key constraint.

Indexes indicate that a column or set of columns should be indexed in the database for faster query performance.
Indexes are defined by a list of one or more columns.
The indexes may be used as an indication to clients that they should consider using these columns in their queries for better performance.

\section{Schema Validation}
The YAML data defining the schema is validated when loaded, ensuring that the schema is correctly defined.
The model comprises a set of Python classes which inherit from Pydantic's \texttt{BaseModel} class and correspond to the different types of schema objects.
The attributes of the classes are defined using Pydantic's \texttt{Field} class, which allows for specifying the data type, default value, and validation rules.
This allows for strict validation of the schema, ensuring that all required fields are present and that the data types are correct.

Additionally, various "business rules" are defined and enforced using Python "validator" functions.
These functions run automatically as part of the validation process.
An example business rule might ensure that table names are unique within a given schema.
Any validation errors which occur will cause the validation process to fail, preventing the schema from being loaded, and an error message will be returned to the user.

\section{Using the Python API}

Felis provides a Python API for working with schema data.
Below is a simple example of working with this API.
\begin{verbatim}
from felis.datamodel import Schema
schema = Schema.from_uri("resource://...")

for table in schema.tables:
    print(f"Table: {table.name}")
    for column in table.columns:
        print(f"  Column: {column.name}")
\end{verbatim}
The basic functionality consists of a set of classes representing the schema objects (\texttt{Schema}, \texttt{Table}, \texttt{Column}, etc.).
Extension modules provide capabilities for generating DDL statements or loading the data into a TAP\_SCHEMA database.
Future extensions are planned to support additional functionality, such as converting tabular data to other formats or performing database migrations.

\section{Felis Data Types}

Felis defines its own system of data types, which are mapped to target database types.
Some commonly used Felis data types include \texttt{boolean}, \texttt{int}, \texttt{char}, and \texttt{string}.
For instance, the string datatype will generate a VARCHAR column in PostgreSQL and MySQL.
Additionally, these data types have corresponding VOTable types used for generating the TAP\_SCHEMA representation.

\section{DDL Generation}
Felis supports generating DDL statements for several different open source database engines, including MySQL and PostgreSQL.
Information about the target database is provided by a runtime-configurable database engine URL.

Felis may also populate a TAP\_SCHEMA database by generating a set of insert statements representing the schema data, which are then executed to populate the database.

\section{Felis in the Rubin Observatory}

The Science Data Model (SDM) Schemas of the Rubin Observatory \citep[e.g.,][]{LSE-163} are defined using the Felis format.
These represent a set of public-facing data catalogs used by the Rubin Science Platform (RSP) to provide access to the data.
The files are managed in a Git repository and versioned as a set with Git tags.
Changes to the schemas are validated in GitHub workflows, ensuring that they are well-defined, with any errors flagged to the user.
An online schema browser is available for viewing the latest versions of the SDM Schemas, as well as the history of changes \footnote{\url{https://sdm-schemas.lsst.io}}.
Rubin Observatory's SIAv2 service \citep{P920_adassxxxiv} uses Felis to define the SIAv2 data model and ensure that the resultant VOTable is conformant.

\section{Future Work}

Felis is under active development, with several new features planned.
Converting tabular data to other formats, such as astropy tables \citep{2022ApJ...935..167A}, is one of these.
Another planned feature is the generation of data migration scripts for updating databases to new schema versions.

\acknowledgments This material or work is supported in part by the National Science Foundation through Cooperative Agreement AST-1258333 and Cooperative Support Agreement AST1836783 managed by the Association of Universities for Research in Astronomy (AURA), and the Department of Energy under Contract No.\ DE-AC02-76SF00515 with the SLAC National Accelerator Laboratory managed by Stanford University.

\bibliography{C702}

\end{document}